\documentclass[twocolumn,prb,superscriptaddress]{revtex4}

\usepackage{graphicx}
\usepackage{amssymb}
\usepackage{textcomp,times} 

\usepackage{jcp}

\newcommand{\strc}[1]{\textsl{#1}}

\begin{document}

\title{\sffamily Activation and protonation of dinitrogen at the 
                FeMo cofactor of nitrogenase}

\author{Johannes K\"astner}
\affiliation{Institute for Theoretical Physics, Clausthal University
  of Technology, D-38678 Clausthal-Zellerfeld, Germany. Electronic mail:
  Kaestner@mpi-muelheim.mpg.de, Peter.Bloechl@tu-clausthal.de}
\affiliation{present address: Max-Planck-Institute for Coal Research,
  Kaiser Wilhelm Platz 1 D-45470 M\"ulheim an der Ruhr, Germany.}

\author{Sascha Hemmen}
\affiliation{Institute for Theoretical Physics, Clausthal University
  of Technology, D-38678 Clausthal-Zellerfeld, Germany. Electronic mail:
  Kaestner@mpi-muelheim.mpg.de, Peter.Bloechl@tu-clausthal.de}

\author{Peter E. Bl\"ochl}
\affiliation{Institute for Theoretical Physics, Clausthal University
  of Technology, D-38678 Clausthal-Zellerfeld, Germany. Electronic mail:
  Kaestner@mpi-muelheim.mpg.de, Peter.Bloechl@tu-clausthal.de}

\begin{abstract} \noindent
  The protonation of N$_2$ bound to the active center of nitrogenase
  has been investigated using state-of-the-art density-functional
  theory calculations. Dinitrogen in the bridging mode is activated by
  forming two bonds to Fe sites, which results in a reduction of the
  energy for the first hydrogen transfer by 123\,kJ/mol. The axial
  binding mode with open sulfur bridge is less reactive by 30\,kJ/mol
  and the energetic ordering of the axial and bridged binding modes is
  reversed in favor of the bridging dinitrogen during the first
  protonation. Protonation of the central ligand is thermodynamically
  favorable but kinetically hindered. If the central ligand is
  protonated, the proton is transferred to dinitrogen following the
  second protonation. Protonation of dinitrogen at the Mo site does
  not lead to low-energy intermediates.
\end{abstract}

\date{Recieved at J. Chem. Phys. 19 May 2005; accepted 5 July 2005; 
published online 22 August 2005}
\maketitle 

\section{Introduction}

Although dinitrogen is the main part of the atmosphere, it is, in its
molecular form, inaccessible to biological organisms. Its inactivity
is caused by the triple bond---one of the strongest covalent bonds in
nature.  While high pressure and high temperature are required to
convert N$_2$ to NH$_3$ in the industrial Haber-Bosch process,
biological nitrogen fixation achieves the same goal at ambient
conditions. For this purpose nature employs the enzyme nitrogenase,
which is one of the most complex bioinorganic catalysts in nature.

Nitrogenase converts N$_2$ to biologically accessible
ammonia.\cite{bur90,bur96,ead96,chr01} During the reaction,
non-stoichiometric\cite{had69} amounts of hydrogen are produced.
\begin{eqnarray*}
\textrm{N}_2+(6+2x)\textrm{H}^++(6+2x)\textrm{e}^-\rightarrow 
  2\textrm{NH}_3+x\textrm{H}_2.
\end{eqnarray*}
Values for $x$ range from $<$1 (Ref. \onlinecite{riv75}) upwards and
there is an ongoing discussion about the question whether hydrogen is
produced in stoichiometric amounts.\cite{ree00} In the gas phase, the
first protonation is by far the most difficult step in the
reaction. The catalyst has to find a way to dramatically reduce this
barrier.

The active center of the enzyme, shown in Fig.~1, is the FeMo
cofactor, MoFe$_7$S$_9$N$\cdot$homocitrate. The FeMo cofactor is
linked to the protein via two amino acid residues.  The question
arises why nature employs such a large multi-center cluster.

The reaction mechanism has being studied for over 40 years, but the
atomistic mechanism of substrate conversion at the FeMo cofactor still
remains an open issue. In a previous paper, we presented the most
salient features of the reaction cycle as it emerged from our
simulations.\cite{kae05c} In the present paper, we present the results
of state-of-the-art first principles calculations of the most
difficult step in the reaction pathway, the first protonation of N$_2$
bound to the FeMo cofactor.

\begin{figure}[t]
\center
\includegraphics[scale=0.8]{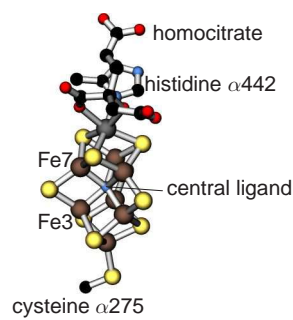}
\caption{The FeMo cofactor of nitrogenase with its homocitrate ligand
and the two residues linking the cofactor to the protein: histidine and
cysteine.\label{fig:rslabel}}
\end{figure}

Nitrogenase consists of two proteins: (1) the molybdenum-iron protein,
which holds the FeMo cofactor (FeMoco) and (2) the iron protein,
which hydrolyzes MgATP and uses the energy obtained to provide the
molybdenum-iron protein with electrons.

The structures of the two proteins were resolved already in
1992.\cite{kim92,kim92a,geo92} Since then increasingly more refined
crystallographic data became available \cite{cha93,pet97,may99} until,
in 2002, a central ligand of the FeMoco, shown in Fig.~1 was
found,\cite{ein02} which was undetected in previous studies.  While
crystallographic studies determined the central ligand to be either a
C, N, or O atom, theoretical studies\cite{hin03,dan03,lov03,hun04}
provide strong support for it to be a nitrogen atom.  It turns out
that the central ligand (N$_\mathrm{x}$) plays a critical role in the
mechanism.\cite{kae05c}

Kinetic studies of the mechanism of biological nitrogen fixation
\cite{low84a,tho84,low84,tho84a} indicate that the rate-limiting step
of the reaction is the dissociation of the two proteins. In each of
these association-dissociation cycles one electron is transferred to
FeMoco. The so-called Thorneley-Lowe scheme provides insight into the
first reduction steps, stating that N$_2$ binds after three or four
electrons have been transferred to the MoFe protein. Theoretical
models\cite{kur01a,lia04} indicate that geometrical changes of the
backbone of the Fe-protein are responsible for using the energy from
MgATP to transfer electrons to the MoFe protein.

Since the appearance of the first crystal structures, nitrogenase has
been subject to numerous theoretical
investigations.\cite{den93,pla94,dan94,dan96,dan97,dan98,mac95,sta96,
sta97,sta98,zho97,sie98,rod99,rod00,rod00a,szi00,szi01,dur01,dur01a,
dur02,dur02a}\\ \cite{bar01,lov01,lov02,lov02a,lov02b,cui03}
Unfortunately, the central ligand was unknown at that time, so that
the conclusions have to be reconsidered. However, several more recent
theoretical studies \cite{hin03,dan03,lov03,vra03,hun04,sch03,kae05c}
have been performed taking the central ligand into account.

A critical parameter for theoretical calculations is the oxidation
state of the cofactor.  By comparing theoretical
results\cite{dan03,lov03,sch03} with experimental observations, a
consensus has been reached that the cofactor in the resting state is
charge neutral, i.e., [MoFe$_7$S$_9$N]$^0$.

The mechanisms for nitrogen fixation proposed up to now may be divided
in two main groups: (1) conversion at Mo and (2) conversion at Fe.
\begin{itemize}
\item The Mo atom has been the target of numerous
  theoretical\cite{pic96,gro98,szi01,dur02,dur02a} studies. A strong
  indication in favor of Mo comes from experiment: an Mo-based model
  complex\cite{yan03} has been found to catalytically reduce
  N$_2$. However, there are also active nitrogenases\cite{ead96} with
  Mo replaced by V or Fe, which indicates that Mo is not essential.
  
\item Numerous proposals have been made for the reduction involving
  the Fe atoms.  (1) The direct way for N$_2$ to bind at Fe atoms is
  head-on binding.\cite{dan97,dan98,rod99,rod00,rod00a,hin04} In this
  mechanism N$_2$ binds to one Fe atom of the central cluster, where
  it is protonated until one and then the second ammonia dissociate.
  The intermediates of the complete cycle have been identified, albeit
  still without the central ligand by Rod and
  co-workers\cite{rod99,rod00,rod00a} Recently, Hinnemann and
  N\o{}rskov\cite{hin04} extended this proposal to the cofactor with
  central ligand. The theoretical predictions are, however, at
  variance with this mechanism, because the calculations show that
  N$_2$ binding is highly endothermic.\cite{hin04} (2) Sellmann and
  co-workers\cite{tho96,sel99,sel00,kir05} suggested an opening of the
  cage in analogy to smaller Fe complexes.  In his model, two
  octahedrally coordinated low-spin Fe atoms positioned in close
  proximity bind dinitrogen between them, where it is reduced. The
  mechanism has been lined out up to the first four protonations.  (3)
  Our own recent calculations\cite{sch03,kae05c} support the view of
  cage opening even though, in our model, the Fe atoms are in a
  high-spin tetrahedral coordination, which points towards a quite
  different chemistry than expected for low-spin octahedral Fe atoms.
  The opening and closing of an SH bridge at the cofactor is critical
  for the activation of N$_2$ and for the dissociation of the second
  ammonia. A critical role is attributed to the central ligand, which
  enables required bond rearrangements. In a similar approach, we were
  able to explain many experimental findings in the conversion of
  acetylene by nitrogenase.\cite{kae05} (4) Another
  proposal\cite{hun04} suggests the opening of a sulfur bridge upon
  coordination of water to an Fe atom, complete protonation of the
  central ligand, and dissociation of ammonia. Then N$_2$ inserts into
  the central cavity of the cofactor, where one nitrogen atom is fully
  protonated and dissociated, which closes the catalytic cycle.  This
  intriguing proposal seems to be in conflict with isotope exchange
  (ESEEM/ENDOR) experiments\cite{lee03} that exclude an exchange of a
  central nitrogen ligand.
\end{itemize}

\section{Calculational Details}

We considered the complete FeMo cofactor with truncated ligands as in
the previous study.\cite{sch03}  The histidine was replaced by
imidazole, the homocitrate ligand by glycolate and the cysteine, bound
to the terminal iron atom by an SH group.

We performed density-functional theory\cite{hoh64,koh65} (DFT)
calculations based on the projector-augmented wave\cite{blo94,blo03}
(PAW) method. The gradient-corrected PBE (Ref. \onlinecite{per96})
functional was used for exchange and correlation.  The PAW method is a
frozen-core all-electron method.  Like other plane-wave-based methods,
the PAW method leads to the occurrence of artificial periodic images
of the structures. This effect was avoided by explicit subtraction of
the electrostatic interaction between them.\cite{blo86} Wave-function
overlap was avoided by choosing the unit cell large enough to keep a
distance of more than 6\,\AA{} between atoms belonging to different
periodic images. We used a plane wave cutoff of 30\,Ry for the
auxiliary wave functions of the PAW method. The following shells were
treated in the frozen-core approximation: Fe [Ne], Mo [Ar3d$^{10}$], S
[Ne], O [He], N [He], and C [He]. The following sets of projector
functions were employed, Fe $2s2p2d$, Mo $2s2p2d$, S $2s2p2d$, O
$2s2p1d$, N $2s2p1d$, C $2s2p1d$, and H $2s1p$, which provides the
number of projector functions per angular momentum magnetic and spin
quantum numbers $m$, $s$ in each main angular momentum channel $\ell$.

Atomic structures were optimized by damped Car-Parrinello\cite{car85}
molecular dynamics with all degrees of freedom relaxed. The
convergence was tested by monitoring if the kinetic temperature
remains below 5\,K during a simulation of 0.05\,ps (200 time steps).
During that simulation, no friction was applied to the atomic motion
and the friction on the wave function dynamics was chosen sufficiently
low to avoid a noticeable effect on the atomic motion.

Transition states were determined by applying a one-dimensional
constraint on the atomic positions. In the present application,
bond-length, angle, and torsion constraints were used. To get a first
upper bound for the barrier, the specific constraint was varied within
1000 molecular-dynamics (MD) steps. If this upper bound is less than
20\,kJ/mol, the barrier will be easily overcome and it has not been
calculated more accurately. In case of a higher estimate, the bond
length was fixed to discrete values around the transition state to
maximize the energy while all unconstrained degrees of freedom were
allowed to relax to minimize the energy. Proof that this approach,
when converged, determines exactly first-order transition states is
given elsewhere.\cite{blo96}

The reaction rates $\Gamma$ can be estimated using $\Gamma=\Gamma_0\,
e^{-E_\mathrm{A}/(k_\mathrm{B}T)}$ from the calculated
activation energy $E_\mathrm{A}$ and a typical vibrational frequency
$\Gamma_0=3\times 10^{13}\,\textrm{s}^{-1}$ corresponding to about
1000\,cm$^{-1}.$

The FeMoco exhibits seven high-spin iron atoms antiferromagnetically
coupled to each other. Many different spin configurations may easily
lead to metastable states in conventional collinear spin-polarized
calculations. Therefore, we used a noncollinear description of the
spin density for our calculations. In a noncollinear description each
one-electron wave function is a two-component spinor wave
function.\cite{san86,kub88,oda98,hob00} This method not only correctly
describes real noncollinear spin states, which occur within the
reaction mechanism, but also avoids artificial barriers between
different spin configurations occurring in collinear calculations. Our
resulting spin distribution is therefore independent of the random
starting conditions. Such a dependence is a common problem of
conventional (collinear) spin-polarized calculations for this system.
We found that the spin ordering depends on subtle changes in the
atomic structure. It also changes between different states of the
reaction mechanism. The spin orderings encountered in our calculations
are given in Fig.~2, where we follow the notation
introduced by Lovell et al.\cite{lov01} noncollinear spin
arrangements have been found, in this study, only for energetically
unfavorable states, which is why we do not specify them further. 

The spin quantum number $S$ is specified alongside with the
corresponding structures in Figs.~3 and 4. A spin with $S=1$, for
example, corresponds to a triplet. Noncollinear states are indicated
by a spin quantum number that differs from half-integer values. This
corresponds, in analogy with the collinear expression, to the absolute
value of the integrated spin density divided by $\hbar$.

\begin{figure}[htbp]
\includegraphics[]{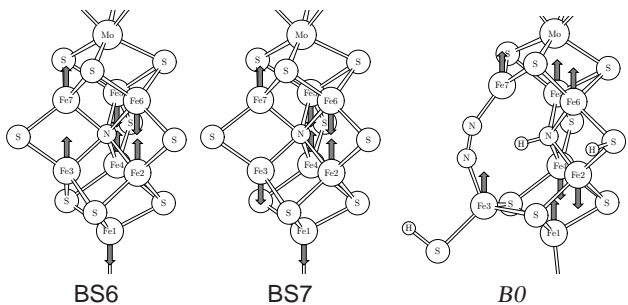}
\caption{The two relevant spin orderings BS6 and BS7 as well as the
  spin ordering of \strc{B0}. BS6 and BS7 differ by a spin flip of
  Fe3. BS7 is the same spin ordering as obtained for the cofactor in
  the resting state M. It is characterized by ferromagnetically
  coupled Fe--Fe and Fe--Mo pairs, which are antiferromagnetically
  aligned relative to each other (Ref. \onlinecite{sch03}). The spin
  ordering only defines the directions of the site spins, not their
  absolute value.
\label{fig:bs}}
\end{figure}


During the reaction, protons and electrons are transferred to the
cofactor and the substrate.  In this work we made the assumption that
electrons and protons are transferred in a correlated way, i.e., that
one proton is transferred with each electron transfer.  This
assumption implies one of the two scenarios: either a reduction of the
cofactor increases the proton affinity such that a proton transfer is
induced, or, if the proton transfer precedes the electron transfer, it
implies that the electron affinity is sufficiently enhanced by the
positive charge next to the cofactor to induce an electron transfer to
the cofactor. This is the main assumption in our work besides the
accuracy of the density functionals and the neglect of the protein
environment. This assumption has been shown to be valid for the
cofactor before the binding of the substrate.\cite{sch03}

The energies of protons and electrons, which are consumed during the
reaction, affect the overall reaction energy. Thus we need to define a
value that reflects their energies in the environment. For protons,
the relevant environment is the proton transfer channel, while for
electrons it is expected to be the P-cluster. The exact energies
cannot be determined by theory alone.

As a consequence of our assumption that the reduction and protonation
occur in a correlated manner, only the sum $\mu_\mathrm{H}$ of the
energies of protons and electrons is relevant for the relative
energies of the intermediates. A range of possible values can be
derived by comparing experimental x-ray and extended x-ray-absorption
fine structure (EXAFS) data with our calculated geometries: we found
indirect evidence for the cofactor being unprotonated in the resting
state, while being protonated in the reduced state.\cite{sch03}
Therefore, $\mu_\mathrm{H}$ is sufficiently high to drive protonation,
that is, $\mu_\mathrm{H}>E($MH$)-E($M$)$. On the other hand, no
protonation occurs under the same conditions, but in the absence of
MgATP. Thus the chemical potential in the absence of MgATP, denoted by
$\mu_\mathrm{H}^\prime$, must be sufficiently low not to drive
protonation, that is, $\mu_\mathrm{H}^\prime<E($MH$)-E($M$)$. As two
MgATP are hydrolyzed in each electron transfer, the difference between
the chemical potentials with and without MgATP is smaller than twice
the energy of hydrolysis of MgATP, that is,
$\mu_\mathrm{H}-\mu_\mathrm{H}'<64.4$\,kJ/mol.\cite{voe02} It is
smaller because a fraction of the energy supplied by MgATP will be
dissipated.  Therefore, we assume the lower bound for
$\mu_\mathrm{H}$, that is, $\mu_\mathrm{H}=E($MH$)-E($M$)$, in our
calculations. This is the most conservative assumption possible. A
less conservative value would make the reactions including protonation
more exothermic.

Previous studies\cite{rod99,rod00,rod00a,hin04,hun04} made a different
choice for $\mu_\mathrm{H}$, namely, $\mu_\mathrm{H}=\frac 1 2
E($H$_2)$. This would be the appropriate choice if the protons and
electrons would be obtained from gaseous hydrogen.  While the
production of gaseous hydrogen 2H$^++2e^-\rightarrow $H$_2$ is
energetically neutral with this choice, this reaction is exothermic by
71\,kJ/mol when using our choice of
$\mu_\mathrm{H}=\frac{1}{2}E($H$_2)+35$\,kJ/mol. Thus our reaction
energies can be translated to the values for H$_2$ reference by adding
35\,kJ/mol per added hydrogen atom. We additionally list the reaction
energies with H$_2$ as the reference energy in parentheses after the
values we obtain with our $\mu_\mathrm{H}$ for all those energies
were influenced by the choice of $\mu_\mathrm{H}$.

In our work we evaluate not only the energetics of the intermediates,
but also the barriers for the transitions. To estimate the barriers
for protonation, we need, however, to specify a proton donor, which
models the proton channel. We used the ammonium ion to mimic the
proton donor. Note, however, that this choice only affects the barriers
but not the relative energies of the intermediates. Our finding that
the barriers for protonations are small and will therefore be overcome
easily indicates that the barriers for protonation are not relevant
for the overall picture.

The notation for the structures is chosen as follows.  The complexes
of dinitrogen with the cofactors are given in letters in alphabetic
order according to the number of proton transfers and in numerals for
their energetic order. A numeral $0$ denotes the ground state for the
selected composition.

\section{Results}

\begin{figure}[htbp]
\includegraphics[]{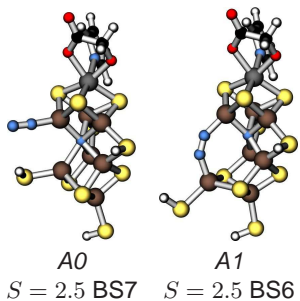}
\caption{N$_2$ binding modes, their spin states, and their spin
ordering (BS state) obtained from DFT.}
\label{fig:n2}
\end{figure}

The most difficult step of the reaction from N$_2$ to NH$_3$ is the
first protonation of dinitrogen. In the reaction in the gas phase we
find that the reaction step N$_2$ + H$^+$ + e$^-\rightarrow$ N$_2$H is
endothermic with 164\,kJ/mol using the choice for $\mu_\mathrm{H}$
described under calculational details (and 199\,kJ/mol with H$_2$ as
reference).  The main goal for nitrogenase must be to lower this
barrier. At the FeMo cofactor, the reaction energy is dramatically
reduced to only 41\,kJ/mol (76\,kJ/mol).  Nevertheless, in the
catalyzed reaction this remains the most endothermic step.

We explored eight different isomers of the N$_2$H adduct bound to the
MH$_2$ state of the cofactor. MH$_2$ represents the resting state (M)
of the cofactor reduced by two electrons and protonated at two of the
sulfur bridges. The isomers are shown in Fig.~4 and will
be discussed in the following.

\begin{figure}[htbp]
\includegraphics[]{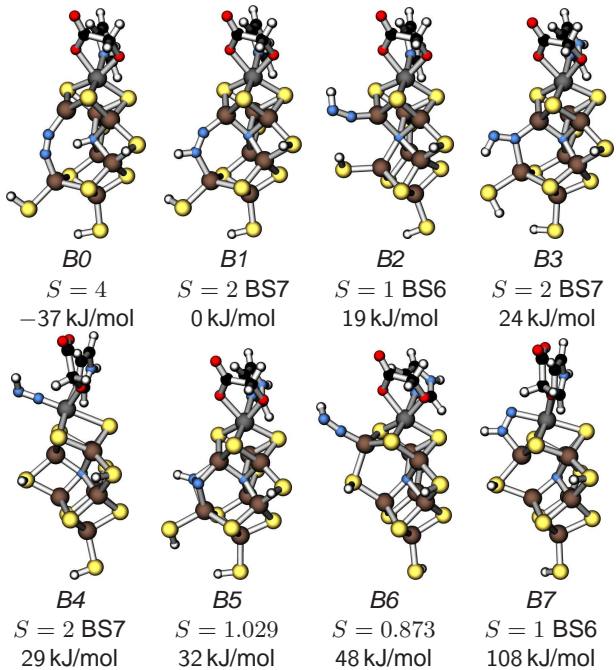}
\caption{Different binding modes of the N$_2$H adduct at the
  FeMoco. The spin state and the spin ordering (BS state) are given,
  as well as the energies, which are given relative to \strc{B1}.}
\label{fig:n2h}
\end{figure}

\subsection{Protonation of the central ligand}
Nitrogen in the bridged binding mode \strc{A1} opens the structure of
the cofactor and leaves space for the coordination at the central
ligand. Protonation of the central ligand leads to the most stable
isomer with one protonated nitrogen, namely \strc{B0}, which is lower
in energy by 37\,kJ/mol than the second most stable configuration, the
protonated dinitrogen bridge \strc{B1}.

There is only limited space for a donor to access the central ligand
in \strc{A1}. Correspondingly, we find a large barrier for protonation
of 79\,kJ/mol from NH$_4^+$, which was used as the model for the
proton donor. This barrier corresponds to a reaction rate of the order
of 0.1--1\,s$^{-1}$, which is substantially lower than the electron
transfer rate. Furthermore, as discussed below, \strc{A1}, the common
starting configuration for both \strc{B0} and \strc{B1} can be
protonated with a negligible barrier at dinitrogen leading to
\strc{B1}. Since the pathways to \strc{B0} and \strc{B1} are in direct
competition, protonation of the central ligand is kinetically
hindered. Therefore, we consider a reaction path via protonation of
the central ligand unlikely.

In \strc{B0}, one iron atom is in a planar threefold coordination
shell, which appears to be an unusual configuration. To explore if
this structure is an artifact of our structural model, we investigated
if this iron atom could form a complex bond to a nearby water
molecule, which would restore a tetrahedral coordination of the Fe
atom.  However, in our calculations including an additional water
molecule, no significant complex bonds between the three-coordinate Fe
atom and a water are formed: for \strc{B0} an Fe--O distance of
2.7\,{\AA} was calculated, which is substantially larger than typical
complex bonds to Fe.

One may ask if the solvent effects affect the barrier to access
\strc{B0}. We find that the interaction between an additional H$_2$O
molecule and the triangular Fe atom in the initial state \strc{A1} and
in the transition state is even weaker than in the final state
\strc{B0}, expressed in even longer Fe--O distances. From the absence
of strong interactions we conclude that the large barrier is not
appreciably influenced by the solvent.

\begin{figure}[htbp]
\includegraphics[]{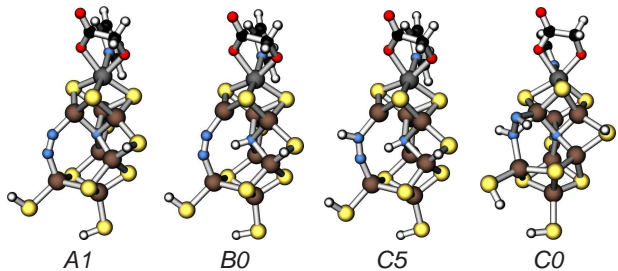}
\caption{Intermediates for the protonation of the central ligand.
\label{fig:brhx}}
\end{figure}

A slow rate of formation does not rule out that the cofactor is
accidentally trapped in this low-energy intermediate. Therefore, we
investigated the subsequent steps starting from \strc{B0}. We find
that after one reduction the next proton attaches to the bridged
dinitrogen, which is similar to the second most stable intermediate
discussed below. However, upon protonation of dinitrogen, the proton
at the central cluster is destabilized and is transferred to the
unprotonated nitrogen atom of dinitrogen. The reaction energy for the
internal proton transfer from the central ligand to the protonated
dinitrogen is $-69$\,kJ/mol. The reaction has a barrier of
20\,kJ/mol. Thus even if the central ligand is protonated, the
reaction mechanism quickly leads back to what we consider the most
likely pathway. Intermediate structures of this side branch of the
reaction cycle are shown in Fig.~5.

\subsection{Protonation of bridged dinitrogen}
The second most stable isomer with one protonated nitrogen is the
bridged binding mode \strc{B1} shown in Fig.~4.

\begin{table}
\caption{Bond distances (\AA), angles (\textdegree), and vibration
numbers (cm$^{-1}$) of the N$_2$ binding modes and free N$_2$. They
are compared to experimental data on N$_2$ and a model complex.
 \label{tab:n2}}
\center
\small
\begin{tabular}{l|ccc|cc}
\hline\hline
  & \multicolumn{3}{c|}{theory} &\multicolumn{2}{c}{experiment} \\
          &  \strc{A0}& \strc{A1} &  N$_2$   &  N$_2$      & 
Model Complex\footnote{Reference \onlinecite{smi01}.} \\
\hline
d(N--N)   &  1.146    &  1.173    & 1.105    &  1.098      &   1.182(5)   \\
d(N--Fe7) &  1.777    &  1.827    &          &             &   1.770(5)   \\
d(N--Fe3) &           &  1.834    &          &             &   1.779(5)   \\
d(SH--Fe3)&  2.298    &  2.349    &          &             &              \\
$\angle$%
(N--N--Fe7)& 176      &  147      &          &             &    180       \\
$\angle$%
(N--N--Fe3)&          &  147      &          &             &    180       \\
$\bar\nu$%
    (N--N)& 1979      &  1792     &  2339    &  2331       &    1778      \\
\hline\hline
\end{tabular}
\normalsize
\end{table}

As shown previously,\cite{sch03} dinitrogen can dock at the FeMo cofactor
MH$_2$ in two configurations with similar energies. They are shown in
Fig.~3. The axial binding mode, denoted by \strc{A0}, is slightly
more stable than the bridged one, denoted by \strc{A1}.

One of the most interesting questions of biological nitrogen fixation
is how dinitrogen is activated for the first protonation. Therefore,
we investigated the two binding modes in some detail.

The most relevant geometric parameters and the N--N stretching vibration
frequencies of dinitrogen bound to the cofactor are given in
Table~\ref{tab:n2} and compared to the experimental data obtained from a model
complex with a N$_2$ bridging two low-coordinated iron atoms.\cite{smi01} The
elongation of the N--N bond as well as the reduction of the vibration
frequency with respect to gaseous N$_2$ are indications of the activation of
the triple bond for the following protonation.  The elongation of the
dinitrogen bond and the reduction of the stretch frequency in \strc{A1}
compares well with the model complex, while \strc{A0} appears to be much less
activated.

The activation of \strc{A1} can be traced back to occupied
N$_2$-$\pi^*$ orbitals as shown in Fig.~6. These orbitals can be seen
as an antisymmetric combination of the Fe--N bonds. Interestingly they
are only dominant in the minority spin direction of the two
neighboring Fe sites Fe3 and Fe7. The origin is that the interaction
with the unoccupied $d$ states split the $\pi^*$ orbital in a bonding
and an antibonding orbital. The bonding orbital, still having $\pi^*$
character but containing the Fe--N bonding contribution, becomes
occupied and is located about 1\,eV below the highest occupied
molecular orbital (HOMO), while the antibonding orbital, having the
Fe--N antibonding contribution, remains unoccupied. The former,
bonding orbital is the relevant frontier orbital for the protonation.

\begin{figure}[htbp]
\center
\includegraphics[]{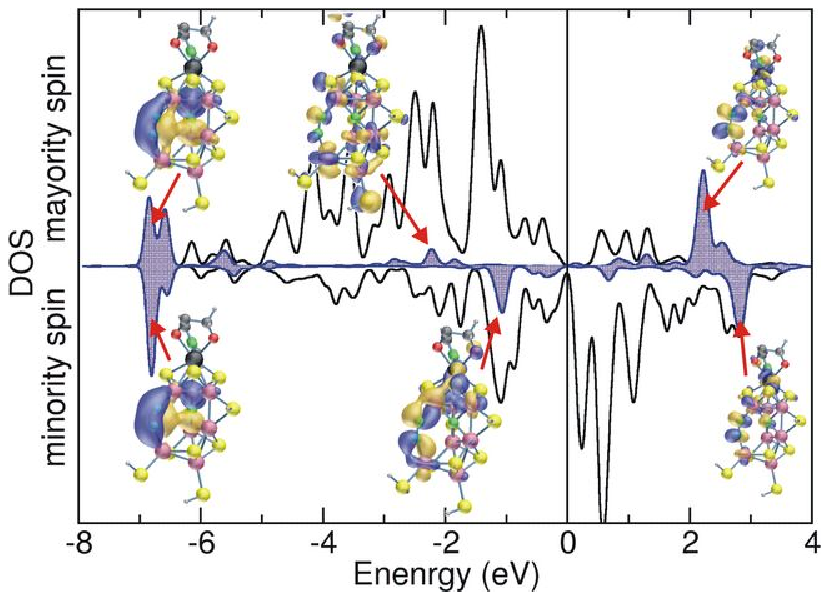}\\
\includegraphics[]{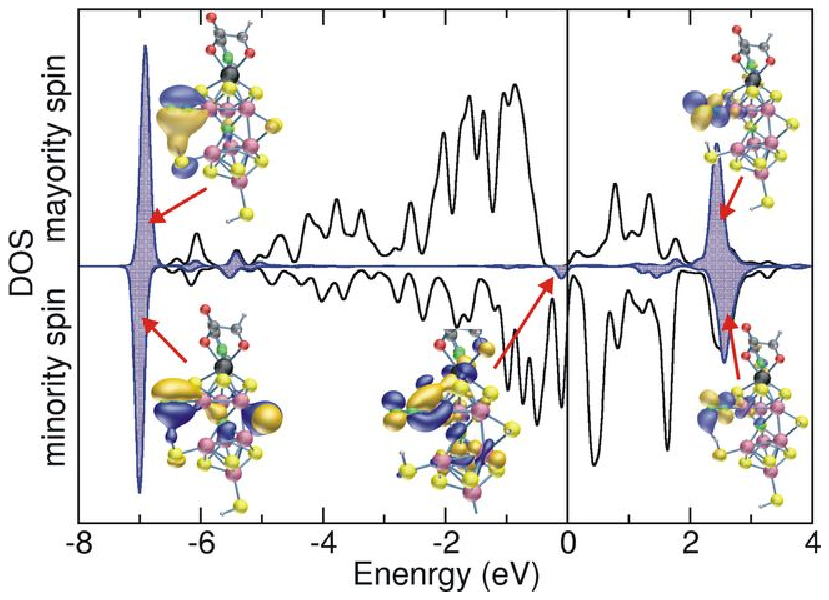}
\caption{Density of states of the bridged (\strc{A1}, top) and the
 axial (\strc{A0}, bottom) binding modes illustrating N$_2$
 activation.  The full line indicates the density of states projected
 onto the $d$ orbitals of the Fe atoms directly bound to N$_2$. The
 shaded curve is the density of states projected onto the $\pi$ and
 $\pi^*$ orbitals of dinitrogen. The insets show the relevant wave
 functions. Note that both states have occupied $\pi^*$ orbitals,
 which is more dominant and lower in energy in the bridged binding
 mode \strc{A1}.  }
\label{fig:dosa0a1}
\end{figure}

The energy to add a hydrogen atom to dinitrogen in the bridged binding
mode \strc{A1} and to obtain \strc{B1} is 41\,kJ/mol (76\,kJ/mol),
substantially less than the 164\,kJ/mol (199\,kJ/mol) of the gas-phase
reaction. \strc{B1} is energetically slightly more favorable by
8\,kJ/mol than protonating the other nitrogen atom of N$_2$.

The protonation of the reduced complex from an ammonium, which mimics
the proton donor in our calculations, proceeds with a negligible
barrier of only 4\,kJ/mol and is exothermic by 63\,kJ/mol. Note,
however, that the calculated protonation energies taken individually
are not as reliable as the reaction energy, as the former depend on
the choice of NH$_4^+$ as the proton donor.  Nevertheless, this
calculation indicates that protonation of the reduced complex at
dinitrogen proceeds without difficulty and much more readily than
protonation at the central ligand.

Interestingly, the spin ordering of the cofactor, as obtained from
DFT, changed from BS6 in \strc{A1} to BS7 in \strc{B1} during the
reduction and protonation. (See Fig.~2 for a definition of the spin
labels.)  This corresponds to a spin flip of the iron atom Fe3, to
which N$_2$ as well as the SH group are bound. In comparison, the FeMo
cofactor in the resting state has the same spin arrangement BS7 as
\strc{B1}. After two protonations and reductions, BS7 of the resting
state is transformed to a noncollinear spin distribution, which
changes to BS7 during dinitrogen binding in the axial mode \strc{A0}
and to BS6 during the conversion to the bridged binding mode
\strc{A1}. These transitions show the importance of allowing spin
flips and noncollinear spins when simulating the reaction. Previous
calculations\cite{rod99,rod00,rod00a,hin04} assumed a fixed spin
ordering during the entire reaction, which may cause errors in the
energy profile.

\subsection{Protonation of axial dinitrogen}
If N$_2$ is axially bound as in \strc{A0} during the first
protonation, the lowest-energy protonation site is the terminal
nitrogen atom. The resulting structure is \strc{B2}, shown in
Fig.~4. The protonation does not induce major structural changes.  The
energy of \strc{B2} is 19\,kJ/mol higher than the energy of \strc{B1},
i.e., the bridged mode. Nevertheless, even though the energy for
protonation of the reduced complex from an ammonium is, compared to
the one of the bridged mode, 30\,kJ/mol smaller, the reaction is still
exothermic with 33\,kJ/mol and proceeds with a negligible barrier of
less than 10\,kJ/mol.

Also in this case, the spin ordering changed during the reduction and
protonation from BS7 in \strc{A0} to BS6 in \strc{B2}.

\subsection{Other binding modes}
Only 5\,kJ/mol above the energy of \strc{B2}, we find the complex
\strc{B3} bridging two Fe atoms with a single nitrogen atom.  This
structural principle is found again later in the reaction
cycle.\cite{kae05c} However, like the axial binding mode, it lies
substantially, that is, by 24\,kJ/mol, higher in energy than the
bridged mode \strc{B1}.

The protonated dinitrogen bound to molybdenum, i.e., \strc{B4}, lies
29\,kJ/mol above \strc{B1}. Together with \strc{B7}, it will be
discussed in Sec.~\ref{sec:mo}, where we discuss the potential role of
molybdenum.

As in \strc{B5}, dinitrogen can also bridge the two Fe atoms with its
axis perpendicular to the Fe--Fe direction, so that both nitrogen
atoms are connected to both Fe atoms. A similar binding mode has
recently been found experimentally for nitrogen bridging two zirconium
centers.\cite{poo04} The energy of this structure, i.e., \strc{B5},
lies 32\,kJ/mol above that of \strc{B1}.

In the relevant intermediate \strc{B6} of the mechanism proposed by
Hinnemann and N{\o}rskov,\cite{hin04} dinitrogen binds axially to one
Fe atom like our structure \strc{B2}. In contrast to \strc{B2},
however, the sulfur bridge is still intact and the bond between this
Fe atom and the central ligand is broken.  This structure is
48\,kJ/mol above \strc{B1} and it is 29\,kJ/mol less stable than the
axial binding mode \strc{B2} discussed previously.

\subsection{Molybdenum as coordination site\label{sec:mo}}

Numerous theoretical \cite{gro98,szi01,dur01,dur01a,dur02,dur02a} and
experimental \cite{bar03,yan03} studies have been performed to
investigate the Mo atom as a possible adsorption site.  For this
reason, we discuss nitrogen coordination to molybdenum here in some
detail.  As shown previously,\cite{sch03} the molybdenum atom is,
according to our simulations, not a favorable adsorption site for
nitrogen: nitrogen bound to molybdenum is higher in energy by
50\,kJ/mol than the one bound to the iron atoms. Nevertheless we
investigate the most favorable pathway for the first protonation via
dinitrogen bound to the Mo site.

As shown previously, nitrogen adsorption at pentacoordinate Mo is
endothermic by 30\,kJ/mol.\cite{sch03} This indicates that, even if
the coordination site is vacant, dinitrogen binds to Mo only for
fleetingly short periods of time.  If protonation proceeds
sufficiently easy so that the proton is transferred during these short
periods, dinitrogen may be stabilized bound to Mo.  However,
protonation leading to \strc{B4}, shown in Fig.~4 is energetically
unfavorable.  \strc{B4} is 29\,kJ/mol higher in energy than \strc{B1},
with N$_2$H bound to Fe atoms.

Durrant\cite{dur02} proposed a transition of the protonated dinitrogen
\strc{B4} into a bridging position between Mo and Fe as in \strc{B7}
shown in Fig.~4. We find \strc{B7} to be 79\,kJ/mol higher in energy
than \strc{B4}.

\section{Discussion}

The following picture emerges from our calculations: dinitrogen exists
in two competing docking modes at the cofactor, the axial mode
\strc{A0} and the bridged mode \strc{A1}.\cite{sch03} They are
separated by a large barrier of 66\,kJ/mol, which, however, is still
sufficiently small to be overcome.\cite{sch03} Before the reduction
and protonation take place, the axially bound dinitrogen is even
slightly more stable by 6\,kJ/mol than the bridged
configuration.\cite{sch03} We assume that axial and bridged modes are
in equilibrium until the proton is transferred.  However, the electron
transfer and the protonation reverse the energetic order between them
so that the axially bound dinitrogen \strc{B2} ends up 19\,kJ/mol
higher in energy than the bridged dinitrogen \strc{B1}.

In the bridged binding mode \strc{A1}, N$_2$ is activated for
accepting a proton by forming bonds to the Fe sites. As a result the
$\pi^*$ orbitals are occupied, which implies that the triple bond is
effectively broken. The occupied $\pi^*$ orbital exposes two
nucleophilic lobes to which a proton can bind.

These occupied $\pi^*$ orbitals are binding combinations of the
dinitrogen $\pi^*$ orbital with the $d$ orbitals of the Fe sites. They
are only occupied in the minority spin direction of the Fe atoms
binding to dinitrogen. This is because only the high-lying $d$
orbitals of the minority spin direction can lead to a stabilization of
the $\pi^*$ orbital; hybridization with the low-lying Fe-$d$ states of
the majority spin direction would shift the $\pi^*$ orbitals further
up, while the corresponding bonding orbital is mostly of Fe character.
Note here that both Fe sites binding to dinitrogen have the same spin
orientation in \strc{A1}.

In the axial binding mode \strc{A0}, the corresponding $\pi^*$ orbital
lies close to the Fermi level and its weight on dinitrogen is
substantially smaller than in the corresponding orbital of the
bridging mode. These differences between the bridged and axial binding
modes are also reflected in the bond-length expansion and the
reduction of the stretch frequency of the dinitrogen bond, and suggest
a smaller activation of the axial binding mode.

Nevertheless, dinitrogen is readily protonated both in the axial and
in the bridged binding mode after the transfer of one electron.
Compared to the gas phase, the reaction energy for the first
protonation is dramatically reduced from 164\,kJ/mol (199\,kJ/mol) for
the gas phase to 41\,kJ/mol (76\,kJ/mol) at the FeMo cofactor in the
bridged mode.

Interestingly, the most stable site for protonation is the central
ligand. The resulting isomer is more stable by 37\,kJ/mol than the one
with protonated dinitrogen. However, this isomer is not considered
relevant because protonation of the central ligand directly competes
with protonation of dinitrogen. While protonation of the central
ligand requires a large barrier of 79\,kJ/mol to be overcome,
protonation of dinitrogen proceeds nearly barrier-less. We attribute
the large barrier to the breaking of the bond between the central
ligand and one of the Fe atoms, which changes its tetrahedral
coordination to a trigonal planar one.

Even if the central ligand is protonated, the second proton is added
to dinitrogen and induces an internal proton transfer from the central
ligand to dinitrogen.\cite{kae05c} Thus even if the first proton binds
to the central ligand, this step neither poisons the catalyst nor does
it lead to an entirely different branch of the reaction cycle. This
side branch may potentially be relevant at small turnover frequencies.

The branch via protonation of the central ligand is reminiscent of the
proposal by Huniar \textit{et al.},\cite{hun04} who propose complete
protonation of the central ligand and its cleavage as ammonia, before
dinitrogen binds to the cofactor. In their study, the central cage is
opened by a water molecule coordinating to one Fe atom. Additional
calculations would be necessary to directly compare our results with
their proposal. However, the latter seems to be in conflict with
isotope exchange (ESEEM/ENDOR) experiments\cite{lee03} that exclude an
exchange of a central nitrogen ligand.

A mechanism via dinitrogen bound to the Mo site is inconsistent with
our calculations. Binding to the Mo atom is substantially more
endothermic than binding to the Fe sites.

Dinitrogen bridging two Fe atoms is part of the proposal by Sellmann
and co-workers\cite{tho96,sel99,sel00,kir05}. Our model differs from
theirs, in that all Fe atoms remain tetrahedrally coordinated and in
the high-spin state. We are not aware of any theoretical
investigations of this system including the cofactor with its central
ligand.

The presence of the central ligand is crucial for the reaction
mechanism of nitrogen fixation at the FeMo cofactor, and may explain
why the mechanism has remained elusive for a long time.  The main
feature of the central ligand is its ability to form a variable number
of bonds to the six Fe atoms.  The central ligand changes its
coordination from sixfold to fivefold and fourfold.  This allows
other ligands such as nitrogen and sulfur to form and cleave bonds to
the Fe atoms without deviations from the preferred tetrahedral
coordination of the latter. Tetrahedral coordination seems to be a
common structural principle to nearly all relevant intermediates of
the reaction cycle.  This is particularly apparent when dinitrogen
binds:\cite{sch03} while the Fe atom to which dinitrogen binds in the
axial binding mode \strc{A0} maintains its coordination shell by
giving up its sulfur bridge, the bridged binding mode \strc{A1} would
result in an unfavorable fivefold coordination of one Fe atom, if the
latter would not give up its bond to the central ligand.

Let us mention here some variants of the reaction steps discussed: our
calculations do not allow to distinguish between Fe3 and Fe7 as
potential docking sites of dinitrogen.  We find that the axial binding
mode \strc{A0} is more stable, that is, by 30\,kJ/mol, on Fe7 than on
Fe3. The bridged mode accessed via the axial binding at Fe7 is more
stable by 14\,kJ/mol than the other variant. We expect that these
differences are sensitive to the environment of the cofactor: due to
the large motion of the sulfur atom, the opening of the SH bridge may
be strongly affected by the shape and the specific interactions of the
cavity. Thus, no conclusive answer regarding the initial binding site
can be given at this point.

The cofactor has an approximate threefold symmetry, which is broken
by the ligands and the protein environment. As long as the protein
environment is not taken into account, as in the present study, the
energetics will proceed similar for all three equivalent orientations.
Nevertheless, the position of the proton channel in the protein
indicates that nitrogen fixation proceeds near the iron atoms Fe3 and Fe7.
Furthermore the cavity in this region provides sufficient space to
accommodate nitrogen bound to an Fe atom.\cite{sch03}

\section{Summary}

We have studied the first protonation of dinitrogen at the cofactor on
the basis of density-functional calculations. We made an effort to
explore the phase space for the reaction without prejudice for one
particular model of the mechanism.  A large number of intermediates
and the barriers between them have been explored and placed into
perspective. While an unambiguous determination of this reaction step is
not yet possible, three possible branches could be identified. One
proceeds via dinitrogen in the bridged configuration, a second one
proceeds via axially bound dinitrogen, and the third proceeds via
protonation of the central ligand. The latter, however, is unlikely to
play a role in the reaction cycle due to kinetic competition with
direct protonation of the bridging dinitrogen.  The activation of
dinitrogen is discussed in detail.

\section*{ACKNOWLEDGMENTS}

We acknowledge support by the HLRN for granting access to their IBM
pSeries 690 supercomputers.  This work has benefited from the
collaboration within the ESF program on ``Electronic Structure
Calculations for Elucidating the Complex Atomistic Behaviour of Solids
and Surfaces.''

\end{document}